\documentclass[10pt,twocolumn,letter]{IEEEtran}
\usepackage{amssymb}
\usepackage{amsmath}
\usepackage{cite}
\usepackage{graphicx}
\usepackage{epstopdf}

\begin{document}

\title{Off-the-grid Two-Dimensional Line Spectral Estimation With Prior Information }
\author{Iman Valiulahi, Hamid Fathi, Sajad Daei, and Farzan Haddadi

}

\maketitle

\begin{abstract}
In this paper, we provide a method to recover off-the-grid frequencies of a signal in two-dimensional (2-D) line spectral estimation. Most of the literature in this field focuses on the case in which the only information is spectral sparsity in a continuous domain and does not consider prior information. However, in many applications such as radar and sonar, one has extra information about the spectrum of the signal of interest. The common way of accommodating prior information is to use weighted atomic norm minimization. We present a new semidefinite program using the theory of positive trigonometric polynomials that incorporate this prior information into 2-D line spectral estimation. Specifically, we assume prior knowledge of 2-D frequency subbands in which signal frequency components are located. Our approach improves the recovery performance compared with the previous work that does not consider prior information. Through numerical experiments, we find out that the amount of this improvement depends on prior information we have about the locations of the true frequencies.
\end{abstract}
\begin{IEEEkeywords}
  Continuous compressed sensing, Prior information, atomic norm,
Line spectral estimation.
\end{IEEEkeywords}

\section{Introduction}
In recent years, there is a growing interest in continuous compressed sensing (CS) as it appears in many signal processing applications such as radar, sonar, array processing, seismology, and remote sensing\cite{rust2006sub},\cite{haardt19952d}. The problem is to recover a signal which is known to be sparse in a continuous domain. In this paper, we investigate line spectral estimation as a special case of continuous CS where the signal of interest is sparse in the continuous frequency domain. Conventional approaches in this field are based on super-resolution methods such as 2-D MUSIC \cite{hua1992estimating}, and Matrix Enhancement Matrix Pencil\,(MEMP)\cite{haardt19952d}. However, these approaches are sensitive to noise and outliers. Moreover, the number of signal frequency components are assumed to be known a priori, while they are not known in practice.\par
In compressed sensing, the goal is to find a sparse vector under the condition that one has some incomplete random linear observations known as $l_0$ problem\cite{candes2006robust},\cite{donoho2006compressed}. Since the $l_0$ problem is NP-hard one can use the best convex relaxation form of $l_0$ problem, which is the $l_1$ norm minimization problem. The required number of measurements for the $l_1$ problem to succeed is $\mathcal{O}(r\log n)$. Unlike the previous approaches, $l_1$ problem does not need the number of non-zero entries of the signal. In continuation of the previous work, Baraniuk in \cite{baraniuk2007compressive} investigated the reconstruction of sparse signals in the discrete fourier transform (DFT) dictionary. In such cases, the classical $l_1$ problem suffers from basis mismatch as it considers the signal is sparse in the discrete domain while in many applications, the signal lies in a continuous dictionary \cite{chi2011sensitivity},\cite{fannjiang2012coherence}.\par
Tang et al. in \cite{tang2013compressed} presents a new concept of ``Compressed sensing off-the-grid''. They assumed the frequencies to take arbitrary values in the signal frequency domain and show that the number of necessary measurements is $\mathcal{O}(r\log r\log n)$ for perfect recovery.\par
2-D line spectral estimation based on convex optimization methods fall into two categories: atomic norm minimization methods introduced first by Chi et al.\cite{chi2015compressive} that is an extended version of \cite{tang2013compressed}, and matrix completion methods known as Enhanced Matrix Completion (EMaC)\cite{chen2014robust}.\par
In the previous works in off-the-grid 2-D line spectral estimation, the only available information is spectral sparsity. However, in practice extra information is often available e.g., in radar and sonar applications, targets are often change their locations very slowly. Therefore, prior approximate knowledge of their locations are available. This can translate to a prior inexact information about the locations of sparse coefficients in the fourier domain when we use array processing techniques to localize the targets.
A major question in 2-D line spectral estimation is whether this additional information can improve the recovery performance. In this paper, the prior information is in the form of some 2-D frequency intervals that are expected to contain the true signal frequency components. We exploit this prior knowledge using weighted atomic norm optimization problem. Dual of weighted atomic norm is converted to a semidefinite program using the well-known theory of positive trigonometric polynomials. We experimentally find out that the more precise is the prior information, the better we can estimate the signal frequencies.

\section{System model}
Consider a data matrix $\mathbf{X}\in\mathbb{C}^{n_1\times n_2}$. Without loss of generality assume that $n_1=n_2=n$. Each element of the matrix $\mathbf{X}$ can be written as a mixture of $r$ exponentials as follows:
\begin{align}
x_\mathbf{k}=\frac{1}{n}\sum_{i=1}^{r} d_ie^{j2\pi \mathbf{f}_i^T\mathbf{k}},
\end{align}
where $d_i=|d_i|e^{j\phi_i}$ is an arbitrary complex amplitude of each exponential with $\phi_i \in [0,2\pi)$, $J=\{0, . . . , n-1\}\times\{0, . . . ,n-1\}$ represents all of indices of the signal, $\mathbf{k}\,=\,(k_1,k_2)\, \in\, J$ and each frequency component
$\mathbf{f}=(f_{1i},f_{2i})$ $\in\,\,[0,1]\times[0,1]\,,\, i=1,..., r$.
The frequency atoms are defined as:
\begin{align}
\mathbf{a}(f_{1i} ):=\frac{1}{\sqrt{n}}[1,...,y_i^{n-1}]^T,\nonumber\\
\mathbf{a}(f_{2i} ):=\frac{1}{\sqrt{n}}[1,...,z_i^{n-1}]^T,
\end{align}
where$\,\,y_i=e^{j2\pi f_{1i}} \,and\,\, z_i=e^{j2\pi f_{2i}}$. The matrix $\mathbf{X}$ in (1) can be written in the following form:
\begin{align}
\mathbf{X}=\mathbf{Y}\,\mathbf{D}\,\mathbf{Z}^{T},
\end{align}
where:
\begin{align}
\mathbf{Y}:=[\mathbf{a}(f_{11} ),...,\mathbf{a}(f_{1r} )]\,\in\, \mathbb{C}^{n\times r},
\end{align}
\begin{align}
\mathbf{Z}:=[\mathbf{a}(f_{21} ),...,\mathbf{a}(f_{2r} )]\,\in\, \mathbb{C}^{n\times r},
\end{align}
and
\begin{align}
\mathbf{D}:=\mathrm{diag} ([d_1,...,d_r])=\mathrm{diag} \,(\mathbf{d})\,\in \,\mathbb{C}^{r\times r}.
\end{align}
Notice that $\mathbf{x}=\mathrm{vec}(\mathbf{X}^T) \,\in\, \mathbb{C}^{\,n^2 }$ is the vectorized form of $\mathbf{X}$ that can be written as follows:
\begin{align}
\mathbf{x}=(\mathbf{Y}\otimes\mathbf{Z})\,\mathbf{d}=\sum_{i=1}^{r} d_i\, \mathbf{a}(f_{1i} )\otimes \mathbf{\mathbf{a}}(f_{2i} ) \nonumber
\\= \sum_{i=1}^{r} d_i \,\mathbf{c}(f_i )= \sum_{i=1}^{r} |d_i| \,\mathbf{c}(f_i,\phi_i ),
\end{align}
where $\otimes$ represents the kronecker product and
$\mathbf{c}(f_i,\phi_i)=\mathbf{c}(f_{1i},f_{2i})e^{j\phi_i}\triangleq\mathbf{\mathbf{a}}(f_{1i} )\otimes \mathbf{a}(f_{2i})e^{j\phi_i}\,\in \mathbb{C}^{\,n^2}$. By the above definition it is easy to see that $ \,\|\mathbf{c}(f_i)\|_2\,$=1.\par
In many linear inverse problems, the goal is to minimize a structure inducing norm with the constraint of observations \cite{chandrasekaran2012convex}. The unit ball of the structure inducing norm is the convex hull of a special atomic set. For instance, convex hull of the standard basis vectors and their outer product as rank-one matrices form the unit ball of the $l_1$ norm and nuclear norm, respectively. We will use the following special atomic norm defined as:
\begin{align}
\|\mathbf{x}\|_{\mathcal A}\,:=\underset{\mathbf{f}_i\in [0,1)\times[0,1),d_i\in \mathbb{C}}{\inf}\Big\{\sum_{i}|d_i|\,\Big{|}\,\,\mathbf{x}=\sum_{i}d_i\mathbf{c}_i(\mathbf{f}_i)\Big\},
\end{align}
where the atomic set $\mathcal{A}$ is given by:
\begin{align}
\mathcal A:={\{\mathbf{c}(\mathbf{f})\,|\,\mathbf{f}\in\,[0,1)\times\,[0,1) \}}\nonumber.
\end{align}
A common way for recovering a sparse signal with partial observations is the atomic norm optimization problem which is defined as:
\begin{align}
\underset{\mathbf{x}^*\in ~\mathbb{C}^{n^2}}{\min}~\|\mathbf{x}^* \|_\mathcal A~~\mathrm{subject\,\, to} ~~\mathcal P_T (\mathbf{x}^* )=\mathcal P_T (\mathbf{x}),
\end{align}
where $T$ is the set of observed entries, and $\mathcal{P}_T (\mathbf{x})$ is a uniform sampling operator of $\mathbf{x}\in\mathbb{C}^{n^2}$.\par
It is often convenient to recover $\mathbf{x}$ indirectly by first finding the true frequencies using the dual problem of (9) and then the problem reduces to a linear system of equations that can be easily solved. The dual problem of (9) is as follows:
\begin{align}
&\underset{\mathbf{q}\in \mathbb{C}^{n^2},\mathbf{Q}\in\mathbb{C}^{n^2\times n^2}}{\max} \quad \langle \mathbf{x},\mathbf{q}\rangle_\mathbb{R}\nonumber\\
&\mathrm {subject\,\,to }\quad  \delta_\mathbf{k}=\mathrm {tr} {[\Theta_\mathbf{k} Q]},\quad \mathbf{k}\,\in\,\mathcal {H}
\end{align}
\[
\quad\quad\quad\quad\quad\begin{bmatrix}
  Q & \mathbf{q} \\
  \mathbf{q}^* & \mathbf{1}
\end{bmatrix}
\succeq 0\,,\quad\mathbf{q}_{T^c}=0,
\]
where $T^c$ is the complement of the set $T$, $\mathbf{q}$ is the dual variable, $\langle.\,,.\rangle_\mathbb{R}$ denotes the real part of the inner product, $\mathcal{H}$ is a halfspace, $\Theta_k $\,is an elementary Toeplitz matrix with unit valuses on its $k$th diagonal and zeros elsewhere, $\Theta_\mathbf{k}=\Theta_{k_2 }\otimes\Theta_{k_1}$, $\delta_\mathbf{k}=1$ if $\mathbf{k}=(0,0)$, otherwise $\delta_\mathbf{k}=0$, and the semidefinite constraint in (10) imposes $\mathbf{Q}\succeq0$. Since the slater's condition is satisfied in the primal problem (9), strong duality holds between (10) and (9).\par
Regarding the fact that the signal frequency components are the peaks of the following dual polynomial with unit modulus, one can estimate the true frequencies \cite{tang2013compressed},\cite{xu2014precise}:
\begin{align}
Q(\mathbf{f})=\,\sum_{\mathbf{k}\in j}^{}q_\mathbf{k} e^{-j2\pi \mathbf{f}^T \mathbf{k}}.
\end{align}\par
In many applications, there exists some prior information about the parameters of the signal. Specifically, in this paper, the prior information is assumed in the signal spectrum. More precisely, assume that there exist $p$ subbands in the frequency domain of the signal which with probabilities $\{P_F(\mathbf{f}_{P_i})\}_{i=1}^p$ are expected to have the true frequency components (Fig.1). Our approach is to use weighted atomic norm problem to exploit prior knowledge in this problem which is defined as follows:
\begin{align}
&\underset{\mathbf{x}^*}{\min}~~\|\mathbf{x}^* \|_{w\mathcal A}\nonumber\\
&\mathrm{subject~~to}~~\mathcal P_T (\mathbf{x}^* )=\mathcal {P}_T (\mathbf{x}),
\end{align}
with $\|\mathbf{x}\|_{w\mathcal{A}}$ defined as:
\begin{align}
\|\mathbf{x}\|_{w\mathcal{A}}\,:=\underset{f_i\in [0,1)\times[0,1),d_i\in \mathbb C}{\inf}\,\,\Bigg\{\sum_{i}w_i|d_i|\,\Big|\,\,\mathbf{x}=\sum_{i}d_ic_i(f_i)\Bigg\},
\end{align}
where $\mathbf{w}=\{w_1,. . ., w_s\}$ is the vector of weights. Now we define $w(\mathbf{f})=w_1=. \,. \,.=w_s$ as a 2-D piecewise constant function associated with $\mathbf{w}$ in each subband. The value of $w(\mathbf{f}_{P_i})$ in subband $P_i$ is $\frac{1}{P_F(\mathbf{f}_{P_i})}$.\par
Although, in contrast to weighted $l_1$ minimization, there does not exist SDP formulation as in \cite[Equation 2.6]{tang2013compressed} for $(12)$\cite{khajehnejad2009weighted}, however, the dual problem of weighted atomic norm minimization can be used to recover the signal frequency components. The dual problem associated with $(12)$ is given by:
\begin{align}
&~\underset{\mathbf{q}}{\max}~\langle \mathbf{q}, \mathbf{x}\rangle_\mathbb{R}\nonumber \\
&~\mathrm{subject~to}~\|\mathbf{q}\|_{w\mathcal A}^*\leq1,~\nonumber\\
&~\mathbf{q}_{T^c}=0,
\end{align}
where $\|\mathbf{q}\|_{w\mathcal A}^*$\,is the dual norm of $\|\mathbf{q}\|_{w\mathcal A}$ which is defined as:
\begin{align}
&\|\mathbf{q}\|_{w\mathcal A}^*\,=\underset{\|\hat{\mathbf{x}}\|_{w\mathcal A}\leq1}{\sup}{\langle \mathbf{q}, \hat{\mathbf{x}}\rangle_\mathbb{R} }\nonumber\\
&~~=\underset{\phi\in[0,2\pi],\mathbf{f}\,\in\,[0,1)\times[0,1)}{\sup}{\Big\langle \mathbf{q}, e^{j\phi}{\mathbf{c}(\mathbf{f})}\frac{1}{w(\mathbf{f})}\Big\rangle_\mathbb{R}}\nonumber\\
&~~=\underset{\mathbf{f}\,\in\,[0,1)\times[0,1)}{\sup}{\Big|\Big\langle \mathbf{q}, {\mathbf{c}(\mathbf{f})}\frac{1}{w(\mathbf{f})}\Big\rangle\Big|}.
\end{align}
\begin{figure}[t]
\centering
\includegraphics[width = 0.35 \textwidth]{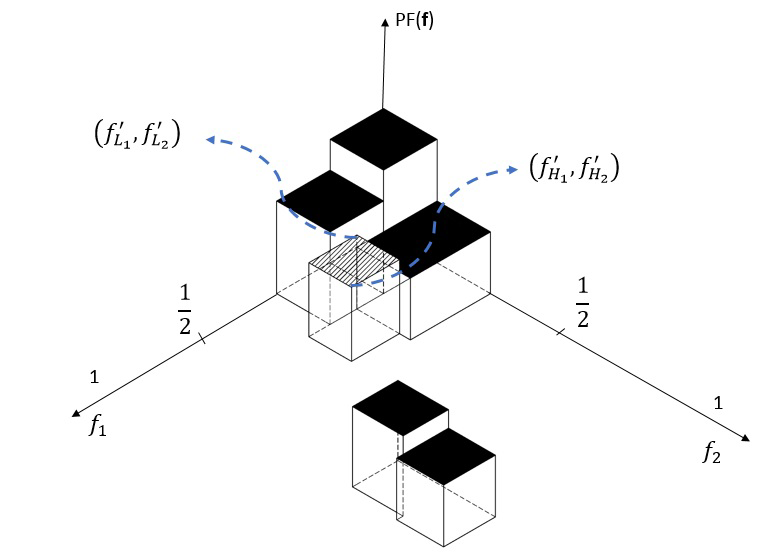}
\caption{Prior distribution of frequency components. The signal spectrum has been normalized to $[0,1)\times[0,1)$} \label{fig:pa}
\end{figure}
Replacing $(15)$ in $(14)$ gives:
\begin{align}
&~\underset{\mathbf{q}}{\max}~\langle \mathbf{q},\mathbf{x} \rangle_\mathbb{R}\nonumber\\
&~\text{subject to}~{\sup_{\mathbf{f}\in[0,1)\times[0,1)}}{|\langle \mathbf{q}, \mathbf{c}(\mathbf{f})\frac{1}{w(\mathbf{f})}\rangle|}\leq1,\nonumber\\
&~\mathbf{q}_{T^c}=0.
\end{align}
Without any prior knowledge $w(\mathbf{f})=1~:~\forall \mathbf{f}\in[0,1)\times[0,1)$ , problem $(10)$ is a special case of the problem $(16)$.\par
Regarding our prior information, problem $(16)$ converts to the following problem with $p$ inequality constraints:
\begin{align}
&~~\underset{\mathbf{q}}{\max}~\langle \mathbf{q},\mathbf{x}\rangle_\mathbb{R}\nonumber\\
&~~\text{subject to}~~{\sup_{\mathbf{f}\in P_i}}{|\langle \mathbf{q}, \mathbf{c}(\mathbf{f})\rangle|}\,\leq w(\mathbf{f}_{P_i}),~i=1,..., p,\nonumber\\
&~~\mathbf{q}_{T^c}=0.
\end{align}
The above relation generalizes the dual problem of 1-D weighted atomic norm in \cite{mishra2015spectral} to 2-D.\par
The following conditions and the positive trigonometric polynomial theory \cite{dumitrescu2007positive} are necessary to convert the inequality constraints (17) to linear matrix inequalities. Let $[f'_{L_1},f'_{H_1}]\times[f'_{L_2},f'_{H_2}] $\ a single frequency band (the hatched subband in Fig. $1$).
\begin{align}
&\bigg \{
\begin{array}{rl}
f_{L_i}=f'_{L_i} \quad\,\,\,&: 0 \leq f'_{L_i} \leq 0.5\nonumber\\
f_{H_i}=f'_{H_i} \quad\,\,\,&:  0 \leq f'_{H_i} \leq 0.5\nonumber\\
\end{array}\\
&\bigg \{
\begin{array}{rl}
f_{L_i}=1-f'_{H_i} &: 0.5 < f'_{H_i} < 1\nonumber\\
f_{H_i}=1-f'_{L_i} &:  0.5 < f'_{L_i} < 1\nonumber\\
\end{array}\\
&\bigg \{
\begin{array}{rl}
f_{L_i}=1-f'_{H_i} &: 0 \leq f'_{L_i} \leq 0.5\\
f_{H_i}=0.5 \quad\quad  &:  0.5 < f'_{H_i} \leq 1\\
\end{array}\\\nonumber\\
&~~~~\cos{\omega_i}-\cos{\omega_{H_i}}\geq0,\nonumber\\
&~-\cos{\omega_i}+\cos{\omega_{L_i}}\geq0,
\end{align}
where $\omega_{H_i}=2\pi f_{H_i},\omega_{L_i}=2\pi f_{L_i}$ and \,$i\in\{1,2\}$. Notice that the complement of the domain $[f'_{L_1},f'_{H_1}]\times[f'_{L_2},f'_{H_2}] $ is obtained by changing the sign of the inequalities in $(19)$.\par
It is well-known in the positive trigonometric polynomial theory \cite{dumitrescu2007positive} that if:
\begin{align}
&\sup {|\langle \mathbf{q},\mathbf{c}(\mathbf{f})\rangle|}\le\gamma\nonumber\\
&\forall~ \mathbf{f}\in [f'_{L_1 },f'_{H_1 }]\,\times\,[f'_{L_2 },f'_{H_2}],
\end{align}
then there exist matrices $G_l\succeq0 ,l\in\{1,...,4\}$, such that:
\begin{align}
&\mathcal G_{\mathbf{k},[f'_{L_1 },f'_{H_1 } ]\times[f'_{L_2}\,f'_{H_2} ] }\triangleq \mathrm{tr}[(\Theta_{k_2}\otimes\Theta_{k_1 } ) \mathbf{G_0} ]+\mathcal T_{\mathbf{k},0.5,-\cos{\omega_{H_1}}} (\mathbf{G_1} )\nonumber\\
&+\mathcal T_{\mathbf{k},-0.5,\cos{\omega_{L_1}}} (\mathbf{G_2} )+\mathcal L_{\mathbf{k},0.5,-\cos{\omega_{H_2}}} (\mathbf{G_3} )\nonumber\\
&+\mathcal L_{\mathbf{k},-0.5,\cos{\omega_{L_2}}} (\mathbf{G_4} )=\delta_\mathbf{k},\nonumber\\
&\quad\quad\quad\quad\mathbf{k}\,\in\,\mathcal H,\quad \begin{bmatrix}
        \mathbf{G_{0}} & \frac{1}{\gamma} \mathbf{q}  \\
  \frac{1}{\gamma} \mathbf{q} ^* & 1\\
     \end{bmatrix}\succeq0,
\end{align}
where $\mathbf{G_0}\,\in\,\mathbb C^{n^2\times n^2} $ and
\{$\mathbf{G_i}\}_{i=1}^4\in \mathbb C^{\,(n-1)^2\times(n-1)^2}$. Moreover, $\mathcal T_{\mathbf{k},d_0,d_1}(\mathbf{G_1})$ and $\mathcal L_{\mathbf{k},r_0,r_1}(\mathbf{G_2})$ are defined as:
\begin{align}
&\mathcal T_{\mathbf{k},d_0,d_1}(\mathbf{G_1})\triangleq \mathrm{tr}[(d_1\Theta_{\mathbf{k_2}}\otimes \Theta_{\mathbf{k_1-1}}+d_0\Theta_{\mathbf{k_2}}\otimes\Theta_{\mathbf{k_1}}\nonumber\\
&+d_1\Theta_{\mathbf{k_2}}\otimes\Theta_{\mathbf{k_1+1}})\mathbf{G_1}],
\end{align}
\begin{align}
&\mathcal L_{\mathbf{k},r_0,r_1}(\mathbf{G_2})\triangleq\,\mathrm{tr}[(r_1\Theta_{\mathbf{k_2-1}}\otimes \Theta_{\mathbf{k_1}}+r_0\Theta_{\mathbf{k_2}}\otimes\Theta_{\mathbf{k_1}}\nonumber\\
&+r_1\Theta_{k_2+1}\otimes\Theta_{k_1})\mathbf{G_2}].
\end{align}
The inequality in (21) implies that $\mathbf{G_0}\succeq0$. After applying $(21)$ to each inequality constraint in $(17)$, the problem $(17)$ changes to the following problem:
\begin{align}
&\underset{\mathbf{q},\mathbf{G_{01}},...,\mathbf{G_{4p}}}{\max} \langle \mathbf{q},\mathbf{x}\rangle_\mathbb{R} \nonumber\\
&\text{subject to}~~\mathcal {G}_{{k_i},{\mathbf{f}}}= \delta_{\mathbf{k}_i}~, ~~\mathbf{f}\in p_i,\nonumber \\
&\mathbf{k}_i\in\{1-n,..,n-1\}\times\{1-n,..,n-1\},\nonumber\\
&\begin{bmatrix}
        \mathbf{G_{0i}} & \frac{1}{w_i(\mathbf{f_{p}}_i)} \mathbf{q} \\
  \frac{1}{w_i(\mathbf{f_{p}}_i)}\mathbf{q}^* & 1
     \end{bmatrix}
     \succcurlyeq 0, \quad i=1,...,p~, \nonumber\\
 &\mathbf{q}_{T^c}=0,
\end{align}
where $\{\mathbf{G}_{0j}\}_{j=1}^p\in\mathbb{C}^{n^2\times n^2 }$ and $\{\{\mathbf{G}_{lj}\}_{l=1}^4\}_{j=1}^p\,\in\, \mathbb{C}^{(n-1)^2\times(n-1)^2}$.\\
problem $(24)$ generalizes the semidefinite programming $(10)$ to the case that one has extra information about the locations of true signal frequencies. As our numerical experiments in the next section verifies, incorporation of prior information improves the frequency recovery performance.
\section{Numerical Experiment}
In this section, we investigate performance of the proposed algorithm by numerical experiments. First, we generate a signal with chi-square distributed amplitude (i.e. $d\sim\delta_{0.5}+\chi^2(1)$ ) and uniform distributed phase $\phi\sim \mathcal{U}(0,2\pi)$. We restrict our experiments to two cases. In the former, we assume that $p=2, s=4, n=7$ and generate four frequency components uniformly in $[0,0.2]\times[0,0.2]$ and $[0.5,0.7]\times[0.5,0.7]$ without imposing any minimum separation condition. Measurements are uniform resampling of a set of $49$ samples without replacement. We assume $\omega(f_{p_1})= \omega(f_{p_2})=1$ and solve (24) using CVX\cite{grant2008cvx} to recover the frequencies. It is shown in Fig. $2$ that the recovery performance improves as we incorporate prior information.\\
In the second simulation we assume $n=7, p=2$. Also, all of the measurements are available. We generate four frequencies uniformly in the domain $[0.1,0.4]\times[0.1,0.4]$ without any separation condition and assume there are prior knowledge in the domain and its complementary given by $P_F(\mathbf{f}_{p_1})=4.13$ and $P_F(\mathbf{f}_{p_2})=0.014$, respectively. Using this information, we obtain $w(\mathbf{f}_{p_1})=0.242$ and $w(\mathbf{f}_{p_2})=71.42$. Now, to recover the frequency components with and without prior knowledge, we solve (24) and (10), respectively. The dual polynomial of weighted atomic norm and regular atomic norm are plotted in Figs. 3(a) and 3(b), respectively. Fig. 3 demonstrates that even by observing all samples, regular atomic norm minimization can not recover the true frequencies properly while the proposed approach clearly identifies the frequency components.

\begin{figure}[t]
\centering
\includegraphics[width = 0.35 \textwidth]{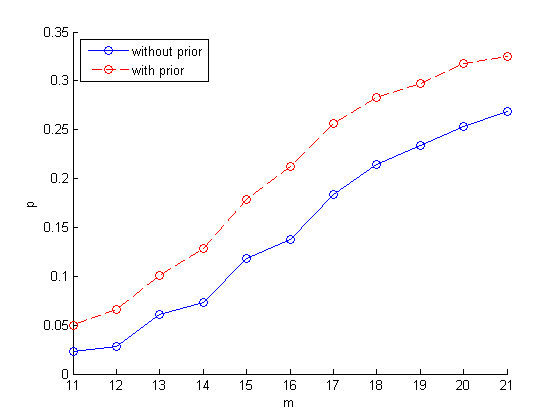}
\caption{Comparison of the proposed method with the case that there is no prior information. Experiments are done with $1000$ trials, $n_1=n_2=7, s=4$ with the additional assumption that the true frequencies are uniformly distributed in $[0,0.2]\times[0,0.2]$ and $[0.5,0.7]\times[0.5,0.7]$ intervals.} \label{fig:pa}
\end{figure}

\begin{figure}[t]
\centering
\includegraphics[width = 0.35 \textwidth]{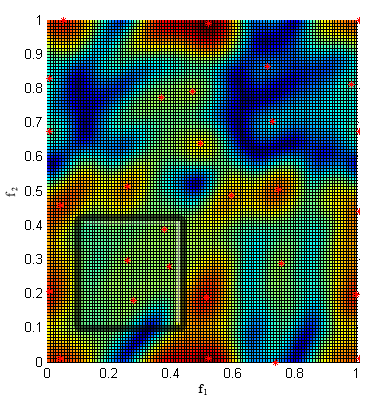}
\begin{center}
{(a)}
\end{center}

\includegraphics[width = 0.35 \textwidth]{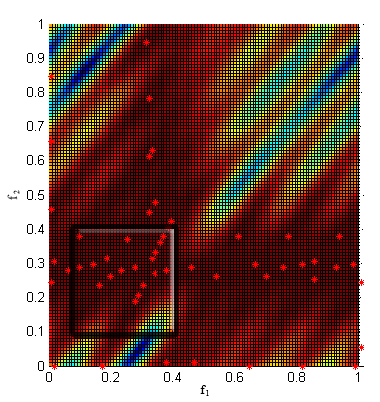}
\begin{center}
{(b)}
\end{center}
\caption{The dual polynomial with $n_1=n_2=7, s=4, m=49$. (a) with prior information about presence of frequencies in $[0.1,0.4]\times[0.1,0.4]$, (b) without prior information.} \label{fig:pa}
\end{figure}
\bibliographystyle{ieeetr}
\bibliography{ref}
\end{document}